\magnification=\magstep 1 \overfullrule=0pt \hfuzz=16pt
\voffset=0.0 true in \vsize=8.8 true in \baselineskip 16pt
\parskip 4pt
\hoffset=0.0 true in
\hsize=6.5 true in
\nopagenumbers
\pageno=1
\footline={\hfil -- {\folio} -- \hfil}

\noindent{}Modern Phys. Lett. {\bf B 16}, 459-465 (2002) {\hfill\hfill} E-print cond-mat/0203039 at www.arxiv.org

\ 

\ 

\

\centerline{{\bf Short-Time Decoherence and Deviation from Pure Quantum States}}

\vskip 0.08 in

\centerline{{Vladimir Privman}}

\vskip 0.04 in

\centerline{\sl Center for Quantum Device Technology, Clarkson
University} \centerline{\sl Potsdam, New York 13699--5820, USA}
\centerline{\sl E-mail: \ privman@clarkson.edu}

\centerline{Received May 8, 2002}

\ 
 
\vskip 0.04 in

\noindent{\bf Abstract: \ }In systems considered for quantum
computing, i.e., for control of quantum dynamics with the goal of
processing information coherently, decoherence and deviation from
pure quantum states, are the main obstacles to error correction.
At low temperatures, usually assumed in quantum computing designs,
some of the accepted approaches to evaluation of relaxation
mechanisms break down. We develop a new formalism for estimation
of decoherence at short times, appropriate for evaluation of
quantum computing architectures.

\vfill\eject

Consider a microscopic quantum system described by the Hamiltonian
$H_S$. This system, $S$, can be a quantum bit (qubit), or several
qubits, controlled externally individually and by switching on and
off pairwise qubit interactions. This qubit paradigm helps to
define the questions in describing how $S$ interacts with the
surrounding macroscopic world. Interactions with the surroundings
depend on the setting. For example, in quantum measurement, the
wavefunction of the system is probed, and part of the process
involves a strong interaction with the measuring device. However,
in most applications, the external interactions are quite weak. In
quantum computing , the aim is to minimize their effect.

Traditionally, interactions with the surroundings have been
modelled by the modes of a ``bath'' $B$, with the mode
Hamiltonians $M_K$,
$$ H_B = \sum_K M_K \; . \eqno(1) $$
\noindent{}The interaction of the bath modes with $S$, will be
modelled by
$$ H_I = \Lambda_S P_B = \Lambda_S \sum_K J_K \; , \eqno(2) $$
\noindent{}where $\Lambda_S$ is a Hermitean operator of $S$,
coupled to the operator $P_B$ of $B$. For a bosonic bath [1,2],
which we use as a prototype,
$$ M_K = \omega_K a^\dagger_K a_K \; , \qquad\qquad J_K = g^*_K a_K + g_K a^\dagger_K \; , \eqno(3) $$
\noindent{}where the ground state is shifted to zero, and we work
in units $\hbar =1$. The total Hamiltonian is then $ H = H_S + H_B
+ H_I $.

Let $\rho (t) $ represent the reduced density matrix of $S$, after
$B$ has been traced over. For large times, the effect of the
environment on an ``idle'' quantum system, i.e., one that is not
otherwise externally controlled, is expected to be thermalization:
this density matrix should approach $ \exp \left(-\beta H_S
\right) \big / {\,{\rm Tr}\,}_S \left[\exp \left(-\beta H_S
\right) \right] $, where $\beta \equiv 1 / k T $. For times $t
\geq 0$, we can consider the degree to which the system has
departed from coherent pure-quantum-state evolution, due to the
interactions and entanglement with the bath. The temperature and
other external parameters needed to characterize $\rho (t) $, are
determined by the properties of $B$, which in turn might interact
with ``the rest of the universe.''

In the eigenbasis of $H_S$, we consider the matrix elements of
$\rho (t)$,
$$ H_S | n \rangle = E_n | n \rangle \; , \qquad\qquad
\rho_{mn} (t) = \langle m | \rho (t) | n \rangle \; . \eqno(4) $$
\noindent{}For large times, we expect the diagonal $\rho_{nn}$ to
approach values $\propto e^{-\beta E_n}$, while the off-diagonal
elements $\rho_{m\neq n} \to 0$, corresponding to thermalization
and decoherence in the energy basis. To establish this, several
assumptions are traditionally made [1-5]. At time $t=0$, it is
assumed that the bath is thermalized, i.e., its modes, $K$, have
density matrices $ \theta_K = e^{-\beta M_K} / {\,{\rm Tr}\,}_K
\left( e^{-\beta M_K} \right)$. The overall density matrix $R$ is
then unentangled,
$$ \Big [ R(t)=\rho(t)\prod_K \theta_K \, \Big ]_{t=0}\; . \eqno(5)$$

Now, a series of assumptions are made, e.g., the Markovian and
secular approximations. The Markovian approximation essentially
assumes that the density matrices of the bath modes are reset to
the thermal ones, on time scales shorter than any dynamical times
of the system interacting with the bath. This amounts to using (5)
for times $t>0$ and is a natural assumption, because each bath
mode is coupled only weakly to the system, whereas it is
``monitored'' by the rest of the universe and kept at temperature
$T$. Such approaches aim at master equations for $\rho_{mn}$ at
large times, consistent with the Golden Rule, thermalization and
decoherence. In variants of these formalisms, two external (to
$S$) times scales are identified. One is the inverse of the upper
cutoff frequency of the bath modes, $1/\omega_D$. Another is the
``thermal'' time $\hbar / kT = \beta $.

There is evidence [3,5,6] that the Markovian-type approximations
are only valid for times large than both these time scales. This
is hardly a limitation at room temperatures. However, for quantum
computing, in semiconductor-heterostructure architectures [7-10],
$T\approx 10-100\,\mu{}$K. The thermal time then becomes
dangerously close to the single-qubit control time even for slower
qubits, based on nuclear spins. We emphasize that not all the
approximation schemes have this problem [5]. We also point out
that quantum computing architectures utilize [7-10] qubits and
modes that couple them, that have large spectral gaps. It is
believed that, especially at low temperatures, spectral gaps slow
down relaxation processes. Thus, qubit levels are considered in
quantum dots, in atoms, in large magnetic fields, and coupled by
highly nondissipative ``quantum'' media [8,10], e.g., the
quantum-Hall two-dimensional electron gas.

At low temperatures, spectral gaps lead to separation of time
scales of the initial decoherence vs.\ later-stage thermalization
and further decoherence driven by energy-conserving processes. One
can then question whether the energy basis is the best to describe
decoherence for times before the thermalizing processes take over.
The issue of the basis has also come up in models of quantum
measurement [11,12], where the eigenbasis of the interaction,
$\Lambda_S$, may be more appropriate. In fact, in quantum
computing applications we really want to retain a pure quantum
state [12,13]. The value of $1-{\,{\rm Tr}\,}_S \left[\rho^2
(t)\right]$, of other measure, may be more indicative than
off-diagonal matrix elements. Therefore, it is desirable to have
basis-independent expressions for the reduced density operator
$\rho (t)$.

Recently, several groups have reported [6,10,12,14-16] results for
spin decoherence in quantum computing systems. Some have not
invoked all the traditional approximations, Markovian and secular,
etc., or have utilized the spectral gap of the bath modes, to
achieve better reliability of the short-time results. Experimental
efforts are picking up momentum, with the first limited results
available [17] by NMR/ESR techniques. An approach, termed
adiabatic decoherence, has been developed [12] to avoid the
ambiguity of the basis selection and achieve exact solvability
[6,12,14,15]. The price paid was the assumption that $H_S$ is
conserved (a version of the quantum nondemolition assumption),
i.e., $ [H_S,H] = [H_S,\Lambda_S] =0 $. This makes the eigenbasis
of $H_S$ and $\Lambda_S$ the same, but precludes energy exchange,
leaving only relaxation pathways that contribute to decoherence.
Generally, $ \Lambda_S | \gamma \rangle = \lambda_\gamma | \gamma
\rangle $, where the Greek index labels the eigenstates of
$\Lambda_S$; the Roman indices will be used for the energy, (4),
and, capitalized, for the bath modes, (2,3).

The most widely used approximation has been the second-order
perturbative expansion in the interaction, $H_I$. In the present
work, we report a different scheme, valid for short times. It has
several advantages, such as becoming exact in the adiabatic case,
yielding several explicit results, and permitting derivation of
higher-order approximations. We do assume that at $t=0$ the system
and bath modes are not entangled. Our formulation also relies on
that the Hamiltonians are all time-independent, and, therefore,
applies to ``idling'' (possibly interacting) qubits. It is
reasonable to assume that a lower limit on decoherence rate can be
evaluated in such an idling state. The $t=0$ factorization
assumption, shared by all the recent spin-decoherence studies,
represents the expectation that control by short-duration but
large externally applied potentials, will ``reset'' the qubits,
disentangling them from the environment to which they are only
weakly coupled. Thus, it is the qubit system that gets
approximately reset and disentangled from the bath at $t=0$,
rather than the bath is thermalized by the ``rest of the
universe,'' as assumed in Markovian approximation schemes.

The overall density matrix is $ R(t) = e^{-i(H_S+H_B+H_I)t} R(0)
\, e^{i(H_S+H_B+H_I)t} $. The following relation for the
exponential factors will be used as our short-time approximation,
$$ e^{i(H_S+H_B+H_I)t + O(t^3)} = e^{iH_St/2}\, e^{i(H_B+H_I)t}\,
e^{iH_St/2} \; . \eqno(6)$$
\noindent{}It becomes exact for the adiabatic-decoherence case.
Furthermore, $e^{\pm iHt}$ are replaced by three consecutive
time-evolution-type transformations on $R(0)$. Therefore, the
approximate expression for $R(t)$ will have all the desired
positivity properties. Extensions to higher-orders in powers of
$t$ are possible: see [18] for various expressions valid to
$O(t^4)$ and $O(t^5)$ in our nomenclature. In the resulting
approximation to the matrix element,
$$ \rho_{mn}(t) = {\,{\rm Tr}\,}_B \langle m | e^{-iH_St/2}\,
e^{-i(H_B+H_I)t}\, e^{-iH_St/2} R(0) \, e^{iH_St/2}\, e^{i(H_B+H_I)t}\,
e^{iH_St/2} |n
\rangle \; , \eqno(7) $$ \noindent we insert the decomposition of
the unit operator in the $S$ space in terms of the eigenbasis of
$\Lambda_S$ before the second exponential, and in terms of the
eigenbasis of $H_S$ after it, etc.,
$$ \rho_{mn}(t) = \sum_{\gamma\, p\, q\, \delta}
{\,{\rm Tr}\,}_B \Big[\,e^{-iE_m t/2}\langle m |\gamma \rangle
\langle \gamma |p \rangle e^{-i(H_B+\lambda_\gamma P_B)t}\,
e^{-iE_p t/2} \rho_{pq}(0) $$
$$\times\Big( \prod_K \theta_K \Big)
e^{iE_q t/2} \, e^{i(H_B+\lambda_\delta P_B)t} \langle q |\delta
\rangle \langle \delta |n \rangle e^{iE_n t/2}\,\Big] \; .
\eqno(8) $$
\noindent We use (1,2) to write
$$ \rho_{mn}(t) = \sum_{\gamma\, p\, q\, \delta}\Big\{
e^{i(E_q+E_n-E_p-E_m)t/2}\langle m |\gamma \rangle
\langle \gamma |p \rangle \langle q |\delta \rangle \langle \delta |n \rangle \rho_{pq}(0) $$
$$\times
\prod_K {\,{\rm Tr}\,}_K  \Big[ e^{-i(M_K+\lambda_\gamma
J_K)t}\,\theta_K\, e^{i(M_K+\lambda_\delta J_K)t} \Big] \Big\} \;
. \eqno(9) $$ \noindent This expression actually allows rather
straightforward calculations in typical quantum-computing
applications which involve single or few two-state systems. The
overlap brackets are between the eigenstates of $H_S$ (labeled by
$m$, $n$, $p$ and $q$) and of $\Lambda_S$ (labeled by $\gamma$ and
$\delta$). The traces are over the modes of $B$. Since these modes
have identical structure, with $K$-dependent coupling constants,
the calculation needs only be done once, in the space of {\it one
mode}.

For the bosonic bath [2], see (3), with the thermal initial
$\theta_K =  \left( 1 - e^{-\beta \omega_K} \right ) e^{-\beta
\omega_K a^\dagger_K a_K}$, the product of traces in (9) is known
[6,12,14],
$$ \rho_{mn}(t) = \sum_{\gamma\, p\, q\, \delta}\Big\{
e^{i(E_q+E_n-E_p-E_m)t/2}\langle m |\gamma \rangle
\langle \gamma |p \rangle \langle q |\delta \rangle \langle \delta |n \rangle \rho_{pq}(0) $$
$$\times
\exp \Big( - \sum_K { |g_K|^2 \over \omega_K^2 } \Big [
2 \left(\lambda_\gamma - \lambda_\delta \right)^2
\sin^2 {\omega_K t \over 2} \coth {\beta \omega_K \over 2}
 +\, i \left(\lambda_\gamma^2 - \lambda_\delta^2 \right) \left(
\sin \omega_K t - \omega_K t \right) \Big ] \Big) \Big \} \; .
\eqno(10) $$ \noindent The last term in the exponent, linear in
$t$, is the ``renormalization'' of the system energy levels due to
its interaction with the bath modes. It could be removed by adding
the term $H_R= \Lambda_S^2 \sum_K |g_K|^2 / \omega_K$ to the total
Hamiltonian. The non-negative real spectral sums, $B(t)$ and
$C(t)$,
$$ B^2(t) = 8 \sum_K { |g_K|^2 \over \omega_K^2 }
\sin^2 {\omega_K t \over 2} \coth {\beta \omega_K \over 2}
 \; , \eqno(11) $$
$$ C(t) = \sum_K { |g_K|^2 \over \omega_K^2 } \left (
\omega_K t - \sin \omega_K t   \right) \; , \eqno(12) $$
\noindent{}when converted to integrals over the bath mode
frequencies, with the cutoff at $\omega_D$, have been discussed
extensively in the literature [2,6,14], for several choices of the
bath mode density of states and coupling strength $g$ as functions
of the mode frequency. In summary, we get the approximation
$$ \rho_{mn}(t) = \sum_{\gamma\, p\, q\, \delta}\Bigg\{
e^{i(E_q+E_n-E_p-E_m)t/2}\langle m |\gamma \rangle
\langle \gamma |p \rangle \langle q |\delta \rangle \langle \delta |n \rangle \rho_{pq}(0) $$
$$\times
\exp \left[ - {1\over 4}B^2(t) \left(\lambda_\gamma -
\lambda_\delta \right)^2 \,+\, i C(t) \left(\lambda_\gamma^2 -
\lambda_\delta^2 \right) \right] \Bigg \} \; , \eqno(13) $$
\noindent{}which is exact for all times in the adiabatic case
[6,12,14,15,19], and has the properties of a density operator.

A basis-independent representation for $\rho (t)$ is obtained by
using $ \sqrt{\pi} \exp [ - B^2 (\Delta\lambda)^2 /4] =
\int_{-\infty}^\infty \! dy \, e^{-y^2} \exp [ i y B
(\Delta\lambda)]$. Exponential factors in (13) can then be
reproduced by applying operators on the wavefunctions, and the
sums carried out,
$$ \sqrt{\pi} \rho =  \int\!
dy \, e^{-y^2} e^{-iH_St/2}
\,e^{i[yB(t)\Lambda_S+C(t)\Lambda_S^2]} \,e^{-iH_St/2} \,\rho(0)
\,e^{iH_St/2} \,e^{-i[yB(t) \Lambda_S+C(t)\Lambda_S^2]}
\,e^{iH_St/2} \; . \eqno(14) $$ \noindent Decoherence is explicit
in (14): if $\rho(0)$ is a projection operator
$|\psi_0\rangle\langle\psi_0|$ (a pure state), then $\rho(t>0)$ is
obviously a mixture (integral over $y$) of projectors
$|\psi(y,t)\rangle\langle\psi(y,t)|$, where $\psi(y,t)=
e^{-iH_St/2} \,e^{i[yB(t)\Lambda_S+C(t)\Lambda_S^2]}
\,e^{-iH_St/2} \,\psi_0$.

As an application, let us consider the case of $H_S$ proportional
to the Pauli matrix $\sigma_z$, e.g., spin-$1/2$ in magnetic
field, and $\Lambda_S = \sigma_x$, with the proportionality
constant absorbed in the couplings $g_K$ in (3). We study the
deviation of the state of a spin-$1/2$ qubit, initially in the
energy eigenstate $|\uparrow\,\rangle$ or $|\downarrow\,\rangle$,
from pure state, by calculating ${\,{\rm Tr}\,}_S \, [\rho^2
(t)]$. For a two-by-two density matrix, this trace can vary from 1
for pure quantum states to $1/2$ for maximally mixed states. With
$\rho (0) = |\uparrow\,\rangle \langle \,\uparrow |$ or
$|\downarrow\,\rangle \langle \,\downarrow |$,
$$ {\,{\rm Tr}\,}_S \, [\rho^2 (t)] = {1\over 2} \left[1+
e^{-2B^2(t)}\right] \; . \eqno(15)$$ \noindent As the time
increases, the function $B^2(t)$ grows monotonically from zero
[2,6,12,14]. For Ohmic dissipation, $B^2(t)$ increases
quadratically for short times $t < O(1/\omega_D)$, then
logarithmically for $O(1/\omega_D) < t < O(\hbar /kT)$, and
linearly for $t>O(\hbar/kT)$. (For other bath models, it need not
diverge to infinity at large times.) This calculation illustrates
that the present approximation can yield reasonable results for
short and even intermediate times.

In summary, we have derived short-time approximations, (13,14),
for the density matrix or its energy-basis matrix elements, for
the case of the bosonic heat bath with initially fully thermalized
modes. Other baths can be studied by using (9).

The author acknowledges useful discussions with Professors Yurii
Bychkov and Tsofar Maniv. This research was supported by the
National Science Foundation, grants DMR-0121146 and ECS-0102500,
and by the National Security Agency and Advanced Research and
Development Activity under Army Research Office contract
DAAD-19-99-1-0342.

\vfill\eject

\centerline{\bf References}{\frenchspacing

\noindent{}{\bf  1. } R.P. Feynman and A.R. Hibbs, {\it Quantum
Mechanics and Path Integrals\/} (McGraw-Hill, NY, 1965); G.W.
Ford, M. Kac and P. Mazur, J. Math. Phys. {\bf 6}, 504 (1965);
A.O. Caldeira and A.J. Leggett, Phys. Rev. Lett. {\bf 46}, 211
(1981), Physica {\bf 121A}, 587 (1983); S. Chakravarty and A.J.
Leggett, Phys. Rev. Lett. {\bf 52}, 5 (1984).{\hfill\break}
\noindent{}{\bf  2. } A.J. Leggett, S. Chakravarty, A.T. Dorsey,
M.P.A. Fisher and W. Zwerger, Rev. Mod. Phys. {\bf 59}, 1
(1987).{\hfill\break} \noindent{}{\bf  3. } N.G. van Kampen, {\it
Stochastic Processes in Physics and Chemistry\/} (North-Holland,
Amsterdam, 2001).{\hfill\break} \noindent{}{\bf  4. } W.H.
Louisell, {\it Quantum Statistical Properties of Radiation\/}
(Wiley, NY, 1973); K. Blum, {\it Density Matrix Theory and
Applications\/} (Plenum, NY, 1996); A. Abragam, {\it The
Principles of Nuclear Magnetism\/} (Clarendon Press,
1983).{\hfill\break} \noindent{}{\bf  5. } Review: H. Grabert, P.
Schramm and G.-L. Ingold, Phys. Rep. {\bf 168}, 115
(1988).{\hfill\break} \noindent{}{\bf  6. } N.G. van Kampen, J.
Stat. Phys. {\bf 78}, 299 (1995).{\hfill\break} \noindent{}{\bf 7.
} D. Loss and D.P. DiVincenzo, Phys. Rev. {\bf A57}, 120
(1998).{\hfill\break} \noindent{}{\bf  8. } V. Privman, I.D.
Vagner and G. Kventsel, Phys. Lett. {\bf A239}, 141
(1998).{\hfill\break} \noindent{}{\bf  9. } B.E. Kane, Nature {\bf
393}, 133 (1998); A. Imamoglu, D.D. Awschalom, G. Burkard, D.P.
DiVincenzo, D. Loss, M. Sherwin and A. Small, Phys. Rev. Lett.
{\bf 83}, 4204 (1999); R. Vrijen, E. Yablonovitch, K. Wang, H.W.
Jiang, A. Balandin, V. Roychowdhury, T. Mor and D.P. DiVincenzo,
Phys. Rev. {\bf A62}, 012306 (2000); S. Bandyopadhyay, Phys. Rev.
{\bf B61}, 13813 (2000).{\hfill\break} \noindent{}{\bf  10. } D.
Mozyrsky, V. Privman and M.L. Glasser, Phys. Rev. Lett. {\bf 86},
5112 (2001).{\hfill\break} \noindent{}{\bf  11. } W.G. Unruh and
W.H. Zurek, Phys. Rev. {\bf D40}, 1071 (1989); W.H. Zurek, S.
Habib and J.P. Paz, Phys. Rev. Lett. {\bf 70}, 1187 (1993); A.O.
Caldeira and A.J. Leggett, Ann. Phys. {\bf 149}, 374 (1983); L.
Mandel and E. Wolf, {\it Optical Coherence and Quantum Optics\/}
(Cambridge University Press, 1995).{\hfill\break} \noindent{}{\bf
12. } D. Mozyrsky and V. Privman, J. Stat. Phys. {\bf 91}, 787
(1998).{\hfill\break} \noindent{}{\bf  13. } P.W. Shor, in {\it
Proc. 37th Annual Symp. Found. Comp. Sci.}, p. 56 (IEEE Comp. Sci.
Soc. Press, Los Alamitos, CA, 1996); D. Aharonov and M. Ben-Or,
quant-ph/9611025, quant-ph/9906129; A. Steane, Phys. Rev. Lett.
{\bf 78}, 2252 (1997); E. Knill and R. Laflamme, Phys. Rev. {\bf
A55}, 900 (1997); D. Gottesman, Phys. Rev. {\bf A57}, 127 (1998);
J. Preskill, Proc. Royal Soc. London {\bf A454}, 385
(1998).{\hfill\break} \noindent{}{\bf  14. } G.M. Palma, K.A.
Suominen and A.K. Ekert, Proc. Royal Soc. London {\bf A452}, 567
(1996).{\hfill\break} \noindent{}{\bf  15. } J. Shao, M.-L. Ge and
H. Cheng, Phys. Rev. {\bf E53}, 1243 (1996); I.S. Tupitsyn, N.V.
Prokof'ev, P.C.E. Stamp, Int. J. Mod. Phys. {\bf B11}, 2901
(1997).{\hfill\break} \noindent{}{\bf  16. } N.V. Prokof'ev and
P.C.E. Stamp, Rep. Prog. Phys. {\bf 63}, 669 (2000); J. Ankerhold
and H. Grabert, Phys. Rev. {\bf E61}, 3450 (2000); T. Maniv, Y.A.
Bychkov, I.D. Vagner and P. Wyder, Phys. Rev. {\bf B64}, 193306
(2001); A.M. Dyugaev, I.D. Vagner and P. Wyder, cond-mat/0005005;
D. Mozyrsky, S. Kogan and G.P. Berman, cond-mat/0112135; A.V.
Khaetskii, D. Loss and L. Glazman, cond-mat/0201303; I.A.
Merkulov, A.L. Efros and M. Rosen, cond-mat/0202271; D. Mozyrsky,
V. Privman and I.D. Vagner, Phys. Rev. {\bf B63}, 085313
(2001).{\hfill\break} \noindent{}{\bf  17. } J. Zhang, Z. Lu, L.
Shan and Z. Deng, quant-ph/0202146; E. Yablonovitch, private
communication.{\hfill\break} \noindent{}{\bf 18. } A.T. Sornborger
and E.D. Stewart, Phys. Rev. {\bf A60}, 1956 (1999).{\hfill\break}
\noindent{}{\bf  19. } Some models of quantum measurement evaluate
decoherence by setting $H_S=0$. Our approximation then becomes
exact: D. Mozyrsky and V. Privman, Mod. Phys. Lett. {\bf B14}, 303
(2000); D. Braun, F. Haake, W.T. Strunz, Phys. Rev. Lett. {\bf 86}
2913 (2001).}

\bye